\documentclass[amsmath, amssymb, aps, prb, superscriptaddress, longbibliography, preprint, citeautoscript]{revtex4-2}

\usepackage[hmargin=2cm,vmargin=2.5cm]{geometry}
\setcitestyle{super}

\usepackage[pdftex]{color}
\usepackage{verbatim}
\pdfoutput=1
\usepackage{float}
\usepackage[section]{placeins}
\usepackage{booktabs,microtype,afterpage} 
\usepackage{graphics}
\usepackage{amssymb}
\usepackage{bm}
\usepackage{braket}

\usepackage[english]{babel}

\usepackage{ulem} 

\usepackage[colorlinks,plainpages=false,linkcolor=blue,urlcolor=blue,citecolor=blue,pdfpagemode=UseNone,pdfstartview=FitBH]{hyperref}
\definecolor{red}{rgb}{0.85,.1,0}
\definecolor{green}{rgb}{0.0,0.6,0.0}
\definecolor{orange}{rgb}{1,0.5,0}
\usepackage{graphicx}    
\graphicspath{{./}{figure/}}
\DeclareGraphicsExtensions{.eps,.png,.pdf,.jpg}

\begin{document}
\title{Layer-dependent antiferromagnetic Chern and axion insulating states in UOTe}


\author{Sougata Mardanya}
\thanks{Corresponding author}
\affiliation{Department of Physics and Astronomy, Howard University, Washington, DC 20059, USA}

\author{Barun Ghosh}
\affiliation{Department of Condensed Matter and Materials Physics, S. N. Bose National Centre for Basic Sciences, Kolkata 700106, India}

\author{Mengke Liu}
\affiliation{Department of Physics, Harvard University, Cambridge, MA 02138, USA}

\author{Christopher Broyles}
\affiliation{Department of Physics, Washington University in St. Louis, St. Louis, MO 63130, USA}

\author{Junyeong Ahn}
\affiliation{Department of Physics,
The University of Texas at Austin, TX 78712, USA}

\author{Kai Sun}
\affiliation{Department of Physics, University of Michigan, Ann Arbor, MI 48109, USA}

\author{Jennifer E. Hoffman}
\affiliation{Department of Physics, Harvard University, Cambridge, MA 02138, USA}

\author{Sheng Ran}
\affiliation{Department of Physics, Washington University in St. Louis, St. Louis, MO 63130, USA}

\author{Arun Bansil}
\affiliation{Department of Physics, Northeastern University, Boston, Massachusetts 02115, USA}
\affiliation{Quantum Materials and Sensing Institute, Northeastern University, Burlington, MA 01803, USA}

\author{Su-Yang Xu}
\thanks{Corresponding author}
\affiliation{Department of Chemistry and Chemical Biology, Harvard University, Cambridge, MA 02138, USA}

\author{Sugata Chowdhury}
\thanks{Corresponding author}
\affiliation{Department of Physics and Astronomy, Howard University, Washington, DC 20059, USA}

\date{\today}

\begin{abstract}
Magnetic topological insulators have received significant interest due to their dissipationless edge states, which promise advances in energy-efficient electronic transport. However, the magnetic topological insulator state has typically been found in ferromagnets (FMs) that suffer from low magnetic ordering temperatures and stray fields. Identifying an antiferromagnetic topological insulator that exhibits the quantum anomalous Hall effect (QAHE) with a relatively high Néel temperature has been a longstanding challenge. Here, we focus on the recently discovered van der Waals (vdW) antiferromagnet (AFM) UOTe, which not only features a high Néel temperature (\(\sim\)150K) but also exhibits intriguing Kondo interaction and topological characteristics. Our systematic analysis of the layer-dependent topological phases based on \textit{ab} initio computations predicts the two-layer UOTe film to be an ideal 2D AFM Chern insulator in which the Hall conductivity is quantized with a fully compensated spin magnetization. By applying an in-plane strain or electric field, we show how the itinerancy of U-5f electrons can be manipulated to trigger a transition between the nontrivial ($C = 1$) and trivial ($C = 0$) phases. Interestingly, the 3-layer UOTe film is found to have zero charge conductance but it hosts a quantized spin Hall conductivity (SHC) with finite magneto-electric coupling, suggesting the presence of an axion insulator-like state. The unique magnetic structure of UOTe supports a layer-tunable topology in which films with an odd number of layers are axion-like insulators, while films with an even number of layers are Chern insulators, and the bulk material is a Dirac semimetal. Our study offers a new intrinsic AFM materials platform for realizing correlated topological phases for next-generation spintronics applications and fundamental science studies.
\end{abstract}

\maketitle

\newpage

\section{Introduction}
The QAHE is a novel topological phenomenon that can lead to dissipationless electrical conductance from the associated chiral edge states \cite{Haldane_1988_QAH,Nagaosa_2010_QAH,Liu_2016_QAH_review}. So far, it has only been realized in ferromagnetic systems\cite{Liu_2008_2DTI,Chang_2013_QAH_exp_CrTI,Deng_2020_QAH_exp_MBT, Serlin_2020_QAH_exp_GN_HBN,Li_2021_QAH_exp_TMD,Qiao_2010_QAH_graphene,Yu_2012_QAH_MagTI,Garrity_2013_QAH_HA,Chen_2014_QAH_TCI,Ren_2016_QAH_atomic,Otrokov_2019_QAH_MBT,Zhang_2019_QAH_moire,Wu_2019_QAH_twist_TMD,Chen_2020_QAH_GN_HBN,Bultinck_2020_QAH_twist_BG,Shi_2021_QAH_GN_HBN,Li_2022_QAH_ML_TMO}, which often exhibit low Curie temperatures, and the ferromagnetic domains responsible for the QAHE make them vulnerable to stray magnetic fields. Unlike FMs, the AFMs are inherently robust against stray magnetic fields. They usually possess higher Neel temperatures and faster switching speeds, making them better suited for spintronics applications \cite{Jungwirth_2016_AFM,Baltz_2018_AFM,Nvemec_2018_AFMopto, Libor_2018_topoAFM,Lei_2020_AFM_mobility,Rahman_2021_vdwAFM}. In most AFMs, different spin sectors are connected by the combined $PT$ symmetry, where $P$ and $T$ denote inversion and time-reversal symmetries, respectively, which prevents the material from realizing the Chern insulating phase. Therefore, to support the QAHE in an AFM state, one must break the $PT$ symmetry without generating a net magnetization. Although various theoretical proposals address this challenging material design issue \cite{Zhou_2016_afmci,Jiang_2018_afmci,Li_2019_afmci,Du_2020_afmci,Ebrahimkhas_2022_afmci,Lei_2022_afmci,Wu_2023_afmci,Guo_2023_afmci,Liu_2023_afmci,Zhou_2023_afmci,liang_2024_afmci}, a materials platform for hosting the AFM Chern insulating state has not been demonstrated.\\

In this study, we focus on UOTe, a material recently identified as a layered topological AFM. It crystallizes in the space group P4/nmm (No. 129) in which the two uranium atoms occupy inversion-symmetric sites (Wyckoff position 2c) with opposite magnetic moments. Oxygen and tellurium atoms surround each uranium atom in a square antiprismatic arrangement to form groups of quintuple layers with Te-U-O-U-Te stacking. Employing density functional theory (DFT) analysis and the total energy difference method, we determined that the magnetic exchange interaction between the two uranium atoms, mediated by oxygen, is antiferromagnetic, with an exchange parameter J1 = -1.72 meV. In contrast, the exchange parameter for the two nearest-neighbor intralayer uranium atoms is ferromagnetic, with J2 = 1.23 meV. For the bilayer UOTe, our calculations indicate that, among three magnetic configurations, udud is the lowest in energy, followed by uddu, then uudd, with energy differences of $\Delta E_{(udud-uddu)}=-1~meV$ and $\Delta E_{(uddu-uudd)} =-20~meV$. We therefore expect the natural magnetic order for repeating quintuple layers to follow an up-down-up-down (udud) pattern. However, previous neutron experiments detected a peak with a propagation vector k = (0, 0, 1/2) \cite{Broyles_2025_UOTe}, which rules out the udud pattern. Given this, the remaining possibilities are uddu and uudd, with uddu being more favorable energetically. Consequently, we consider the antiferromagnetic (AFM) ground state to consist of two quintuple layers with reversed magnetic orderings (uddu), separated by a van der Waals (vdW) gap, as illustrated in Fig.\ref{fig1}(a). Additionally, the collinear nature of the AFM state is confirmed by the strong easy-axis anisotropy of 2 meV/U-atom along the c-axis.
The observed Neel temperature of the AFM UOTe is $T_N\approx150$K, which is among the highest values reported in vdW AFMs \cite{Rahman_2021_vdwAFM}. \\ 

The bulk electronic structure of UOTe is semimetallic, where the band overlap around the Fermi level occurs only around the $\Gamma$ point, see Fig.~\ref{fig1}(c). Valence bands are mainly derived from Te-p orbitals, while the conduction bands are dominated by U-f and U-d orbitals. An examination of the low-energy spectrum reveals that the band inversion is driven by the strong spin-orbit interaction of U-5f electrons. These inverted bands form a Dirac-like crossing along the $\Gamma-Z$ direction, protected by the rotational symmetry. This unique magnetic structure of UOTe, which hosts nontrivial topology and strongly interacting 5f electron, demands a systematic theoretical study for identifying possible exotic phases. Here, using density functional theory and model analysis, we investigate the electronic and topological properties of a few layer UOTe films.\\

For an even number of layers, the $PT$ symmetry is broken in UOTe. In the two-layer system, for example, the spin-orbit coupling (SOC) induces a topological band inversion around the Fermi energy to yield chiral edge states and the emergence of the AFM Chern insulator with Chern number $C=1$; the presence of QAHE provides further confirmation. External factors like in-plane strains or perpendicular electric fields can transform this Chern insulator into a trivial insulator ($C=0$) by causing a double-band inversion. For an odd number of layers, even though the ${PT}$ symmetry is preserved, $P$ and $T$ are both individually broken, resulting in an axion-insulator-like phase with layer-tunable magneto-electric coupling \cite{Varnava_2018_AI,Liu_2020_AI,Ahn_2022_AI,Zhuo_2023_AI,Qiu_2024_AI}. We predict the presence of spin Hall conductivity (SHC) in the 3-layer system \cite{sinova_2015_SHE,matthes_2016_SHE,qiao_2018_SHE}.

\section{Results}
\subsection{AFM Chern insulating phase in 2-layer system}
The realization of the AFM Chern insulator phase lies in the breaking of the combined $PT$ symmetry. In bulk UOTe, although time-reversal symmetry is absent, the $PT$ symmetry remains protected via an effective time-reversal symmetry created by combining $T$ with the translation symmetry ($t_z$). In the 2-layer UOTe, this effective time-reversal symmetry is absent, which breaks the overall $PT$ symmetry while maintaining a net zero spin magnetization in the AFM configuration. To investigate the effect of this symmetry breaking on topology, we calculated the electronic structure using the HSE06 hybrid functional (Fig.~\ref{fig2}(b)). A global gap with band inversion is evident at the Fermi energy. A calculation of the Wilson loop over the 2D  Brillouin zone (BZ), see Supporting Information (SI), shows that the system is a topologically nontrivial insulator with $C=1$. The nontrivial bulk topology is further confirmed by the presence of a chiral edge state along the [100] direction inside the bandgap, which connects the conduction and valence bands across the Fermi energy (Fig.~\ref{fig2}(c)). The unidirectional flow of current at the edge results in a quantized Hall conductivity $e^2/h$ within the gap regime (red curve in Fig.~\ref{fig2}(d)).\\ 

The quantized anomalous Hall effect typically appears in systems with uncompensated magnetism, where the majority contribution comes from spin magnetization. Since the spin magnetization in UOTe is zero in each AFM layer, a small contribution will come only from the orbital magnetization, which we calculated using the  expression:
\begin{equation}
\mathbf{M}_{\mathrm{orb}}=\frac{e}{2 \hbar c} \operatorname{Im} \sum_n \int \frac{d^3 k}{(2 \pi)^3} f_{n \mathbf{k}}\left\langle\partial_{\mathbf{k}} u_{n \mathbf{k}}\right| \times\left(H_{\mathbf{k}}+E_{n \mathbf{k}}-2 \mu\right)\left|\partial_{\mathbf{k}} u_{n \mathbf{k}}\right\rangle
\end{equation}
Here, $E_{n \mathbf{k}}$ and $\left|u_{n \mathbf{k}}\right\rangle$ are the eigenvalues and eigenvectors of the Hamiltonian $H_{\mathbf{k}}$, $\mu$ is the chemical potential and $f_{n \mathbf{k}}$ is the Fermi occupation factor. The total orbital magnetization can be decomposed into two components: $\mathbf{M}_{\mathrm{trivial}}$ and $\mathbf{M}_{\mathrm{topological}}$. The trivial component originates from the local circulation of electrons, whereas the topological component arises from circulating currents at the edges \cite{Thonhauser_2005_morb,Lopez_2012_morb_wann}. In two dimensions, the orbital magnetization follows the relation, $\frac{d \mathbf{M}_{\mathrm{orb}}}{d \mu}=\frac{e}{c h} \mathbf{C}$, indicating that inside the gap of a Chern insulator, the magnetization will vary linearly with chemical potential $\mu$ \cite{Thonhauser_2011_morb}. In 2-layer UOTe, the variation of $\mathbf{M}_{\mathrm{orb}}$ with $\mu$ is presented in Fig.~\ref{fig2}(d) (blue line), supporting this predicted linear relationship. At the Fermi energy, the total orbital magnetization amounts to 0.08 $\mu_B$.\\

We note that the AFM configuration of UOTe is stable in the thin-film limit. A temperature-dependent in-plane resistance measurement for a 25 nm flake of UOTe shows a Neel temperature of 140 K (Fig. 2(e)), with little change from the bulk value of about 150K. The intrinsic AFM Chern insulating phase with such a high Neel temperature makes the 2-layer UOTe viable for practical applications, surpassing previously reported candidate materials (Fig. 2(f)).\\

\subsection{Tuning the topological phase via strains and external fields}
We now explore the effects of perturbations on the topological state of 2-layer UOTe. The 5f electrons in the actinides possess features of both core and valence electrons, and depending on the oxidation state and chemical environment, they can exhibit localized or itinerant behavior. In UOTe, the itinerant nature of the 5f electrons drives the topology of the system. Our analysis of the electronic states around the band inversion point shows that the primary contribution to these states comes from the U-5f  and Te-p orbitals. Therefore, by tuning the itinerancy of U-5f electrons, one can expect to change the topological state of the system. Such an exercise not only offers insights into the robustness of the Chern insulating phase but also aids in understanding how external parameters can be used to switch between trivial and nontrivial states applications. We will illustrate this effect below by taking strain and electric field as external control parameters. In this connection, note that when we reduce the itinerancy of the 5f electron via strain or an external electric field, the system transitions to a topologically trivial state. Note that the trivial phase here differs significantly from the putative trivial insulating phases, which involve the absence of boundary states. Here, an additional band inversion appears at the Fermi energy, as shown in the schematic of Fig.~\ref{fig3}(a), which is opposite in character to the intrinsic band inversion responsible for the Chern insulating phase. The two band inversions carry oppositely moving chiral currents, resulting in a topologically trivial system with $C=0$.
An in-depth analysis of the phase transitions driven by changes in external parameters can become computationally quite demanding with the HSE06 functional. Therefore, we employed the more efficient DFT+U approach \cite{hubbardU, Anisimov_1997}, which often captures (qualitatively) reasonably, the changes in the salient features of the energy spectrum. For pristine UOTe, the linear response method gives an estimate of the effective Hubbard $U_{eff}$  of 3.47 eV. Along this line, we modeled the strain effects by adjusting the lattice parameters using the equation $a_{new}=a(1\pm \frac{s}{100})$, where $s$ represents the strain percentage. Since the critical value for the phase transition also depends on the Hubbard U strength, we treat it as another independent variable ranging from 3.5 eV to 4.5 eV, which yielded the topological phase diagram of Fig.~\ref{fig3}(b). At lower $U_{eff}$ values, the system is mostly seen to remain in the Chern insulating phase ($C=1$). However, as we increase the value of $U_{eff}$, the system transitions to a trivial phase ($C=0$) with the outward expansion of the unit cell. The critical strain values at each $U_{eff}$ value are represented by the red color and fitted to identify the phase boundary. The band structures and evolution of the Wannier charge centers for selected U values at this topologically distinct phase are shown in SI.\\

A similar phase transition can also be induced by applying an external electric field perpendicular to the layers. The resulting electric-field gradients in vdW materials can localize electron density. To capture this effect, we adapt the expression of layer-resolved Chern density as formulated in Ref. \cite{Varnava_2018_AI} to obtain the contribution to the Chern number for layer $l$ as: 
\begin{equation}
C_z(l)=\frac{-4 \pi}{A} \operatorname{Im} \frac{1}{N_{\boldsymbol{k}}} \sum_{\boldsymbol{k}} \sum_{v v^{\prime} c} X_{v c \boldsymbol{k}} Y_{v^{\prime} c \boldsymbol{k}}^{\dagger} \rho_{v^{\prime} v \boldsymbol{k}}(l) ,
\end{equation}
where $X_{v c k}=\left\langle\psi_{v k}\right| x\left|\psi_{c k}\right\rangle=\frac{\left\langle\psi_{v k}\right| i \hbar v_x\left|\psi_{c k}\right\rangle}{E_{c k}-E_{v k}}$, and a similar expression holds for $Y_{v c k}$, $\psi_{c k}$. $\psi_{v k}$ are the eigenstates for conduction (c) and valence bands (v), $N_{\boldsymbol{k}}$ is the total number of $\boldsymbol{k}-$points in the 2D BZ and $A$ is the unit cell area. $\rho_{v v^{\prime} \boldsymbol{k}}(l)=\sum_{j \in l} \psi_{v \boldsymbol{k}}^*(j) \psi_{v^{\prime} \boldsymbol{k}}(j)$ is the projection onto layer $l$, where $j$ is the associated orbital. The layer-dependent Chern numbers without and with the external electric field are presented in Fig.~\ref{fig3}(c). Note that without an electric field, each layer gives a positive contribution to the Chern number, with a significant contribution coming from the middle part of the layers. As we apply the electric field, the appearance of the additional band inversion results in both negative and positive contributions from the layers, consequently making the Chern number zero. The spin-resolved band structure in Fig.~\ref{fig3}(d) shows the expected double band-inversion mechanism as an effect of the electric field. The edge spectrum associated with this trivial phase indicates the presence of two edge states with opposite chirality at the Fermi energy.\\

\subsection{Axion insulator phase for odd number of layers and magneto-electric coupling}
In UOTe films with an odd number of layers, the PT symmetry is preserved, resulting in doubly degenerate bands. For the 1-layer film, the conduction and valence bands are separated due to quantum confinement effects with a conventional band gap. In the 3-layer case, however, the conduction and valence bands undergo a band inversion, but symmetry constraints forbid the appearance of topological edge states. Although $C=0$ for both the 1-layer and the 3-layer film enforced by $PT$ symmetry, films with an odd number of layers support an 'axion insulator-like' phase. The presence of inversion or time-reversal symmetry in a bulk axion insulator ensures a quantized axion angle, $\theta=\pi$. In the isotropic case, the axion angle is connected to the magneto-electric coupling as $\alpha=\frac{\theta}{2\pi}\frac{e^2}{h}$, with the magnetization $\textbf{M}=\alpha\textbf{E}$, $\textbf{E}$ being the electric field. To explore the axion insulating phase in odd-layered UOTe films, we calculated $\alpha$ and $\theta$ using Eqs.[\ref{alphas}-\ref{theta}] below
\begin{align}
\label{alphas}
\alpha_{xx} &= {\frac{e^2 }{AL_z}\sum_{m,n}\sum_{\bf k}\frac{f_{nm}}{\varepsilon_{mn}}}\ {\rm Re}[r_{{nm}}^x\braket{\psi_{m\textbf{k}}|-\frac{1}{2}\left(\hat{v}^y\hat{r}^z+\hat{r}^z\hat{v}^y\right)|\psi_{n\textbf{k}}}], \\
\alpha_{yy} &= {\frac{e^2 }{AL_z}\sum_{m,n}\sum_{\bf k}\frac{f_{nm}}{\varepsilon_{mn}}}\ {\rm Re}[r_{{ nm}}^y\braket{\psi_{m\textbf{k}}|\frac{1}{2}\left(\hat{v}^x\hat{r}^z+\hat{r}^z\hat{v}^x\right)|\psi_{n\textbf{k}}}],\\ \nonumber
\alpha_{zz} &= {\frac{e^2 }{AL_z}\sum_{m,n}\sum_{\bf k}\frac{f_{nm}}{\varepsilon_{mn}}\ {\rm Re}[\frac{1}{2}\sum_{p;\varepsilon_p\neq\varepsilon_m}(r_{nm}^zr_{mp}^xv_{pn}^y-r_{np}^zr_{pm}^xv_{mn}^y
+r^z_{nm}r^x_{mn}v^y_{mm}
-(x\leftrightarrow y))}],
\\ 
\end{align}
Here, $m,n$ are the band indices, $\hat{v}$, $\hat{r}$, and $\hat{s}$ are the velocity, position, and spin operators, respectively, and $m_e$ is the electron mass. The $A, L_z$ are the area of the unit cell and the thickness of the sample. Brillouin zone integration is replaced by the ${\bf k}$-summation; $f_{mn}=f_m-f_n$, where $f_n$ is the Fermi Dirac distribution function with energy $\epsilon_n$ at crystal momenta ${\bf k}$. The spin-related last term of the equations above involving $\hat{s}$ represents the spin contribution to $\alpha$, while other terms arise from orbital contributions. We have ignored the contribution to $\alpha$ from the Fermi surface. The Axion angle $\theta$ is computed in terms of $\alpha$ as:
\begin{align}
\theta=\frac{\pi}{3}\frac{e^2}{2h}\sum_{i=x,y,z}\alpha_{ii}
\label{theta}
\end{align}

In pristine 1-layer and 3-layer UOTe, we find a substantial deviation of $\theta$ from the quantized value. For example, in the 1-layer case, $\theta$ is only $\approx 9\%$ of $\pi$, while for the 3-layer film $\theta$ is $\approx 21\%$ of $\pi$. Such a large deviation from the quantized value is consistent with the broken inversion and time-reversal symmetries in the odd-layered films, and it can be understood as a finite-size effect. Variation of $\alpha$ with chemical potential is presented in Fig.~\ref{fig4}(c). For 1-layer, $\alpha$ reduces sharply and changes sign on electron as well as hole doping, while for the 3-layer film, $\alpha$ remains almost constant with hole doping but reduces with electron doping. 

In addition to the axion phase, we consider how the strong SOC-driven band inversion in the 3-layer system might harbor other interesting features, such as the spin Hall effect (SHE), where a traverse spin current appears in response to an electric current. In time-reversal symmetric systems like the two-dimensional topological insulators, the SHC becomes quantized to $\frac{2e^2}{h}$. Although the SHE can be divided into intrinsic and extrinsic parts, the major contribution comes from the intrinsic part, which arises from relativistic effects in the band structure, and it can be calculated from DFT using the Kubo formula:
\begin{equation}
\begin{aligned}
\sigma^{\gamma}_{\alpha \beta}(\omega)= \frac{\hbar}{V N_k^3} \sum_k \sum_n f_{n k} \times \sum_{m \neq n} \frac{2 \operatorname{Im}\left[\langle \psi_{n \boldsymbol{k}}| \hat{j}^{\gamma}_{\alpha}|\psi_{m \boldsymbol{k}}\rangle\langle \psi_{m \boldsymbol{k}}|-e \hat{v}_{\beta}|\psi_{n \boldsymbol{k}}\rangle\right]}{\left(\epsilon_{n k}-\epsilon_{m k}\right)^2-(\hbar \omega+i \eta)^2},
\end{aligned}\end{equation} where, $\alpha, \beta$, and $\gamma$ are cyclic combinations of the Cartesian components $x,y$, and $z$, $\hat{j}^{\gamma}_{\alpha}=\frac{1}{2}\{\frac{\hbar}{2}\hat{\sigma}_{\gamma},\hat{v}_{\alpha}\}$ is the spin-current operator, $\hat{v}_{\alpha} = \frac{1}{\hbar}\frac{\partial H(k)}{\partial k_{\alpha}}$ is the velocity operator, $f_{nk}$ is the Fermi function, $N_k$ is the total number of k-points in the BZ, $V$ is the volume of the unit cell, $n$ and $m$ denote band indices, and $\epsilon_n$ and $\epsilon_m$ are the corresponding band energies. Frequency $\omega$ and the smearing parameter $\eta$ are set to zero for the direct-current clean limit. Note that the 3-layer UOTe film features a large SOC-driven band inversion gap. The analysis of the edge spectrum along [100] also reveals a pair of PT-symmetric edge states inside the gap, which come with opposite chirality (see SI). Although the $PT$ symmetry enforces a zero total charge current at the edge, the SHC is still possible. Using the Kubo formula, we calculate the SHE component $\sigma^x_{xy}$, $\sigma^y_{xy}$ and $\sigma^z_{xy}$ with varying chemical potential, as shown in Fig.~\ref{fig4}(d). The Berry-curvature-like component seen in $\sigma^{\gamma}_{\alpha\beta}$ is obtained by following the recently proposed approach\cite{qiao_2018_SHE} based on a model obtained by using maximally localized Wannier functions and a {$100\times100\times1$ dense k-point grid to sample the BZ. The $\sigma^z_{xy}$ component of SHC forms a quantized plateau at $2 e^2/h$ inside the gap, while the other two component $\sigma^x_{xy}$ and $\sigma^y_{xy}$, are close to zero, suggesting that the edge states of 3-Layer UOTe are fully spin-polarized along the z direction, which is further verified by the spin-polarized edge-state spectrum (see SI). 

\section{Summary}
We have investigated layer-dependent topological properties of UOTe. The absence of $PT$ symmetry in even-layered films is shown to give rise to a Chern insulating phase. Our analysis of a 2-layer film, representative of even-layered films, predicts the presence of a robust AFM Chern insulating phase with negligible net magnetization. We investigate the effects of external perturbations, such as strains and electric fields, on the Chern insulating phase. For odd-layered films, preserved $PT$ symmetry prevents the emergence of the Chern insulating phase. However, since the inversion and time-reversal symmetries are broken, odd-layered films exhibit axion electrodynamics with thickness-dependent magneto-electric coupling. The three-layer system exhibits $PT$-symmetric edge states with opposite chirality and spin, leading to finite SHC. Our study establishes UOTe in the thin-film limit as a rich platform for exploring novel topological quantum phenomena and for developing spintronics applications. 

\section{Methods}
Electronic structure calculations were carried out by employing the first-principles density functional theory (DFT) framework\cite{hohenberg1964inhomogeneous} using the Vienna {\it ab-initio} simulation package (VASP) \cite{kresse1996efficient, kresse1999ultrasoft}. For the bulk case, experimental cell parameters were used, and ionic positions were optimized until the residual force on each ion was less than 10$^{-2}$ eV/\AA $~$, and the stress tensors became negligible. The optimized structure was used to construct slabs of various numbers of layers by placing a vacuum of 10\AA ~between the slabs. Calculations for the 2-layer film are based on the HSE06 hybrid functional. Effects of strains and external electric fields were modeled by introducing an effective on-site Coulomb interaction parameter ($U_{\rm eff}$ on U-f orbitals  \cite{hubbardU, Anisimov_1997}. Spin-orbit interaction was included in all cases. Brillouin zone integrations used a 13$\times$13$\times$6 and 13$\times$13$\times$1, $\Gamma$-centered $k-$mesh \cite{Monkhorst1976} for bulk and slab systems, respectively. An energy cut-off of 500 eV was set for the plane-wave basis set. The total energy tolerance criterion for self-consistency cycles was set at $10^{-8}$ eV. Weak interlayer van der Waal's interactions were treated via the DFT-D3 correction of Grimme with zero-damping function \cite{grimme2010consistent}. Topological properties were calculated by employing material-specific tight-binding model Hamiltonians constructed from atom-centered Wannier functions ~\cite{marzari1997maximally}. Bulk topological character and edge-state spectrum were obtained by using the WannierTools software package~\cite{wanniertool}, while transport simulations use Wannierberri~\cite{wannierberri}. Axion coupling, $\theta$, was computed by using the Wannier function-based models.

\section{Acknowledgements}
We acknowledge fruitful discussions with Igor Mazin. M. L. acknowledges the Harvard Quantum Initiative Postdoctoral Fellowship. The work at Howard University is supported by the U.S. Department of Energy (DOE), Office of Science, Basic Energy Sciences Grant No. DE-SC0022216. The research at Howard University used the resources of Accelerate ACCESS PHYS220127 and PHYS2100073. The work at Northeastern University was supported by the National Science Foundation through the Expand-QISE award NSF-OMA-2329067 and benefited from the resources of Northeastern University's Advanced Scientific Computation Center, the Discovery Cluster, the Massachusetts Technology Collaborative award MTC-22032, and the Quantum Materials and Sensing Institute. The work at Washington University is supported by the National Science Foundation (NSF) Division of Materials Research Award DMR-2236528. C. Broyles acknowledges the NRT LinQ, supported by the NSF under Grant No. 2152221. The work at S. N. Bose National Centre for Basic Sciences (SNBNCBS) was supported by Prime Minister Early Career Research Grant (PM-ECRG) from Anusandhan National Research Foundation (ANRF), file number ANRF/ECRG/2024/003677/PMS, and also benefited from the PARAM-Rudra computational facility at SNBNCBS. 

\section{Author contributions}
S.M., B.G., S.X., and S.C. conceptualized the manuscript. S.M., B.G., A.B., and S.C. performed the DFT calculations and the low-energy model calculations. S.M., B.G., J.A., K.S., S.X., and S.C. did the analyses on the electronic and topological properties. C.B. and S.R. grew the crystal and performed the electric transport measurement. S.M., B.G., M.L., J.E.H., S.X., and S.C. wrote the manuscript with contributions from all authors.

\section{Competing interests}
The author(s) declare no competing interests.

\bibliographystyle{naturemag}
\bibliography{Ref_fixed.bib}

\clearpage\section{Figures}

\begin{figure*}[htb]
\centering
\includegraphics[width=\textwidth]{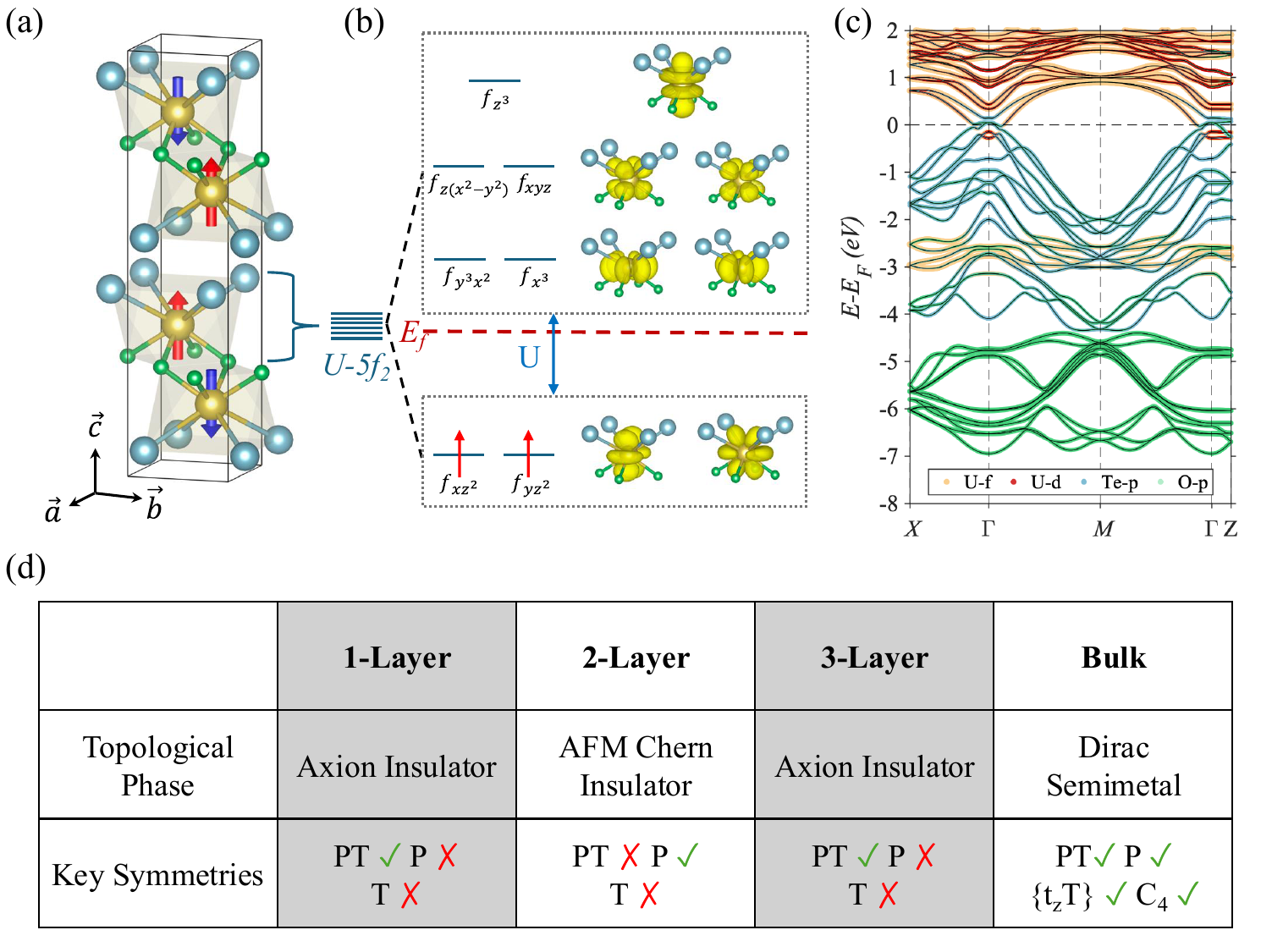}
\caption{\textbf{Crystal structure and layer-dependent topology in UOTe.}(a) Crystal structure and the experimental magnetic configuration of bulk UOTe. Two Uranium atoms in a quintuple layer are connected via two Oxygen atoms, resulting in antiferromagnetic coupling through a superexchange mechanism. (b) Square anti-prismic crystal field splitting of $U-5f$ electrons. (c) Orbital-resolved band structure of bulk UOTe with SOC. The color and size of the markers on the band structure represent different orbitals and their weights, respectively. Band inversion between the $Te-p$ and $U-f$ bands is prominent around the Fermi level. (d) Summary of the layer-dependent topological phases and their symmetries.}
\label{fig1}
\end{figure*}

\begin{figure*}[htb]
\centering
\includegraphics[width=\textwidth]{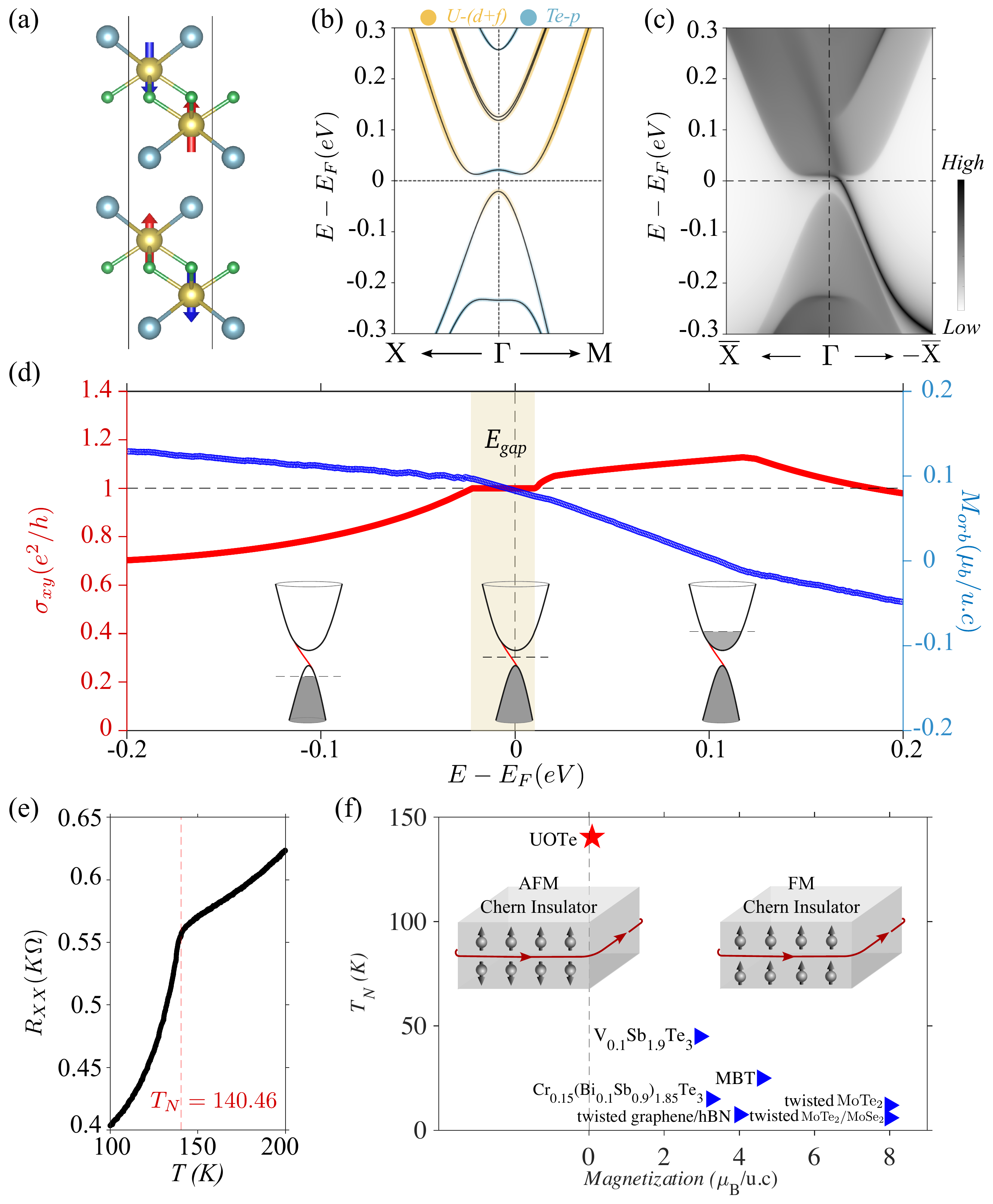}
\caption{(caption next page)}
\label{fig2}
\end{figure*}
\addtocounter{figure}{-1}
\begin{figure*} [t!]
  \caption{\textbf{AFM Chern insulating phase in 2-layer UOTe.} (a) Crystal structure with open boundaries along the c-direction. (b) Orbital-resolved band structure, including the effect of SOC using HSE06 hybrid functional. A clear gap is seen around the Fermi energy, along with a band inversion between the conduction and valence bands. (c) The edge spectrum projected along the [100] direction shows the chiral edge state within the 2D bulk gap. (d) Variation of Hall conductivity (red curve) and orbital magnetization (blue curve) as a function of energy. The circulating edge current results in quantized Hall conductivity and generates a small orbital magnetization within the 2D bulk gap. (e) Temperature-dependent in-plane resistivity for a 25 nm flake shows the discontinuity at 140.46K, indicating the Neel temperature for the AFM transition. (f) Magnetization vs transition temperature phase diagram for comparing the predicted behavior of the 2-layer UOTe film with that of other reported Chern insulators \cite{Tcakaev_2020_magphase,Tschirhart_2021_magphase, Xu_2023_magphase, Mei_2024_magphase},  highlighting the uniqueness of the predicted AFM Chern insulating phase in the 2-layer UOTe film.}
\end{figure*}

\begin{figure*}[htb]
\centering
\includegraphics[width=\textwidth]{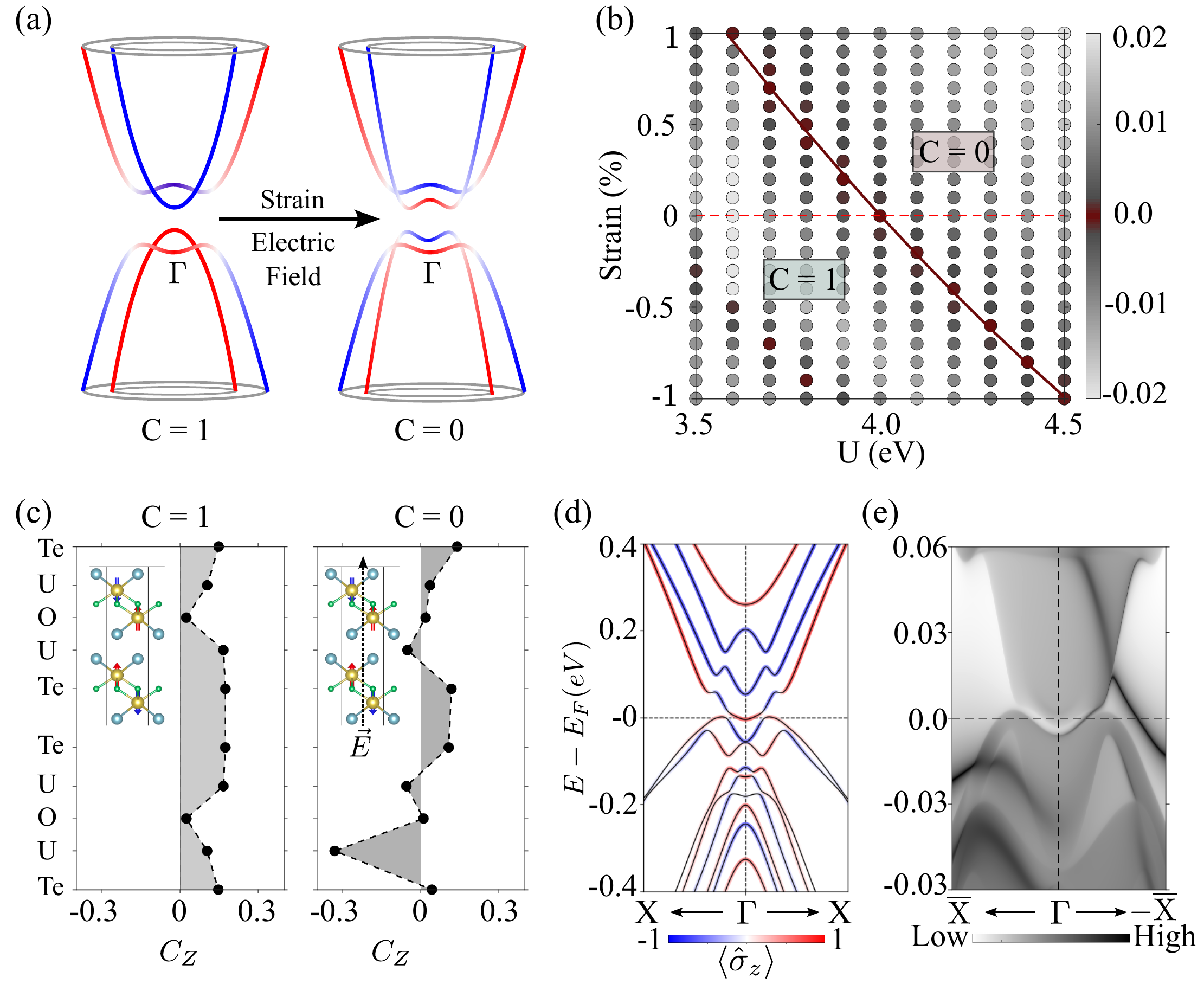}
\caption{\textbf{Effects of strains and electric fields.} (a) A schematic of the double-band- inversion mechanism under external effects such as strains and electric fields to show the transition from the nontrivial (topological) to the trivial phase. (b) $U_{eff}$ vs. bi-axial strain phase diagram, which shows two topologically distinct phases with Chern numbers C=1 and C=0. Intensity of the color represents the size of the bandgap between the highest occupied and lowest unoccupied band at the $\Gamma$ point. Negative values denote inverted band gaps and the presence of a double band inversion. Red line is fitted to the critical points to identify the phase boundary. (c) Layer-dependent Chern numbers capture the effect of the electric field, which changes the topological phase from C=1 to C=0 in the 2-layer UOTe film. (d) $\langle \sigma_z \rangle$-spin-resolved band structure for the 2-layer UOTe film with 0.005 $\mathrm{V/\AA}$ electric field, where a second band inversion can be seen. (e) The edge spectrum projected along the [1-10] direction shows the presence of two oppositely moving chiral states at Fermi energy, which add up to give a topologically trivial phase with C=0.}
\label{fig3}
\end{figure*}

\begin{figure*}[htb]
\centering
\includegraphics[width=0.9\textwidth]{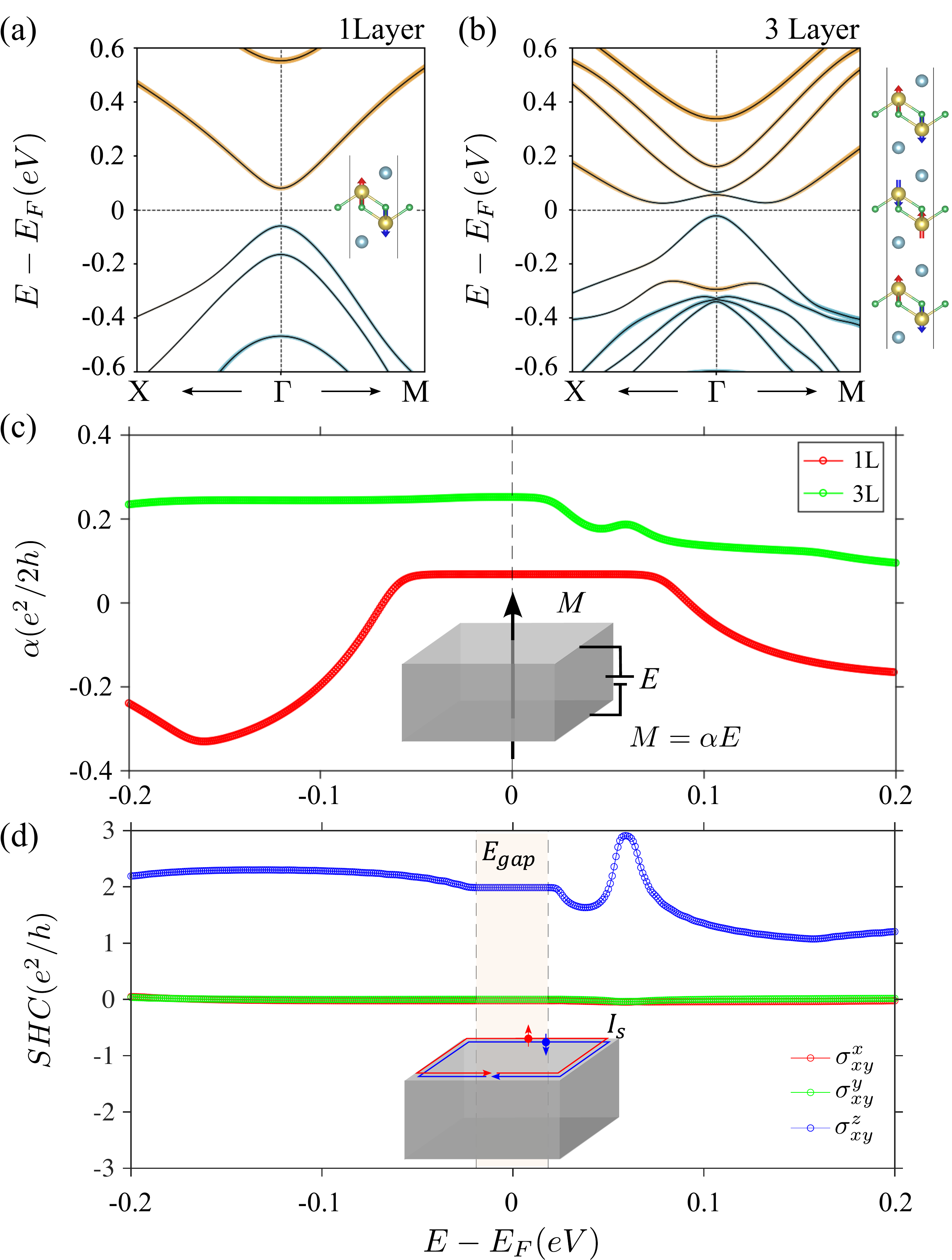}
\caption{(caption next page)}
\label{fig4}
\end{figure*}
\addtocounter{figure}{-1}
\begin{figure*}[t!]
    \caption{\textbf{Magneto-electric coupling and SHC in odd layered UOTe.}(a) Band structure of the 1-layer UOTe film with SOC shows a trivial gap between the valence and conduction bands. (b) Band structure of the 3-layer UOTe film with SOC showing band inversion at the $\Gamma$ point. The crystal structure of the various layers shows that the inherent AFM structure respects the PT symmetry. (c) Variation of magneto-electric coupling with the energy of the 1-layer (red curve) and 3-layer (green curve) films. Magneto-electric coupling is enhanced as the layer thickness increases. (d) SHC for the 3-layer film shows that $\sigma^z_{xy}$ is quantized at $2e^2/h$ inside the gap, suggesting the presence of spin-polarized edge states.}
\end{figure*}

\end{document}